\documentclass[11pt,twoside]{article}

\usepackage{asp2004}
\usepackage{epsf}
\usepackage{graphicx}
\usepackage{lscape}
\usepackage{natbib}

\begin{document}

\title{What Really is the Relativistic Radio-Jet Distance of the Galactic Microquasar GRO~J1655--40?}  

\author{C\'edric Foellmi} 

\affil{European Southern Observatory, 3107 Alonso de Cordova, Casilla 19, Vitacura, Santiago, Chile}  

\affil{Laboratoire d'Astrophysique de l'Observatoire de Grenoble (LAOG), 414 Rue de la Piscine, Domaine Universitaire, 38400 Saint-Martin d'H\`eres, France.}  

\begin{abstract} 
GRO~J1655--40 is a galactic microquasar, i.e. a short-period binary with relativistic radio-jets and where the companion is a black-hole. A little after its discovery in 1994 by $BATSE$, a distance of 3.2$\pm$0.2~kpc has been estimated, and at which the radio jets appeared superluminal. Since that time, this distance of GRO~J1655--40 has been discussed in many studies, often strengthening the value of 3.2~kpc, and used in numerous models. However, recently, \citet{Foellmi-etal-2006b-astroph} used new VLT-UVES and {\it published} photometric data to  show that GRO~J1655--40 must be closer than 1.7~kpc and that the accuracy of the main distance estimators for this source can be questioned. Still, the details on how really the distance of 3.2~kpc has been build was not fully clear. It is the purpose of this article to show that while the {\it upper} limit to the distance is rather firm, the lower limit is not. We draw some conclusions about the new understanding we have of GRO~J1655--40, and finally present a new and promising method that can be used to determine the distance of GRO~J1655--40, and maybe many faint and embedded stars.
\end{abstract}


\section{Introduction}

Microquasars are short-period stellar binaries in which one of the component is a compact object, i.e. a neutron star or a black-hole. The presence of such object is causing (sometimes persistent) strong radio jets and X-ray flares. In some cases, the relativistic jets appear to be superluminal, i.e. having an apparent speed larger than the speed of light. Microquasars are galactic laboratories of high-energy phenomena, and they must be seen as part of a large paradigm where AGNs, microquasars and possibly gamma-ray bursts (GRBs) share similar physics \citep{Mirabel-2004}. The main advantages of microquasars over their large-scale parents, is their short timescales, allowing much more detailed dynamical studies. 

GRO~J1655--40 (a.k.a. Nova Sco 94) has been discovered as a Soft X-ray Transient (SXT) on July 27, 1994 with \emph{BATSE} on board the \emph{Compton Gamma Ray Observatory} \citep{Zhang-etal-1994}. Strong evidence that the compact object in GRO~J1655--40 ~is a black hole was presented by \citet{Bailyn-etal-1995b} \citep[see also][]{Balucinska-Church-2001}. They obtained through spectroscopy a mass function of the secondary (i.e. the absolute minimum mass of the compact object) of $f$ = 3.16$\pm$0.15 $M_{\odot}$; that has been later updated by \citet{Orosz-Bailyn-1997} to $f$ = 3.24$\pm$0.09 $M_{\odot}$. This value is above the theoretical maximum mass of a neutron star: $\sim$3 $M_{\odot}$ \citep{Arnett-Bowers-1977} \citep[but see also][ for a discussion on the dependance of this maximum mass with the equation of state]{Nauenberg-Chapline-1973,Prakash-etal-1988,vanKerkwijk-etal-1995}. Later, \citet{Shahbaz-etal-1999} revised down the mass function below this limit ($f$ = 2.73$\pm$0.09 $M_{\odot}$), but published the masses of the compact object and the secondary star: 5.5--7.9 $M_{\odot}$ and 1.7--3.3 $M_{\odot}$ respectively thanks to the inclination angle, clearly confirming the presence of a black-hole. The jets of GRO~J1655--40 were observed in radio, and given a distance of about 3 kpc, they appeared to be superluminal \citep[see e.g.][]{Rees-1966}: 1.5$\pm$0.4~$c$ \citep{Tingay-etal-1995}, 1.05~$c$ \citep{Hjellming-Rupen-1995}, where $c$ is the speed of light. 

All studies until the most recent ones \citep[e.g.][]{Barret-etal-1996,Regos-etal-1998,vanderHooft-etal-1998,Phillips-etal-1999,Shahbaz-etal-1999,Kuulkers-etal-2000,Soria-etal-2000,Greene-etal-2001,Combi-etal-2001,Buxton-Vennes-2001,Yamaoka-etal-2001,Gierlinski-etal-2001,Kubota-etal-2001,Remillard-etal-2002,Kong-etal-2002,Kobayashi-etal-2003,Stevens-etal-2003,Willems-etal-2005,Brocksopp-etal-2006} usually quote the distance of 3.2$\pm$0.2~kpc determined by \citet{Hjellming-Rupen-1995}, although \citet{Mirabel-etal-2002} pointed out that a distance of about 1~kpc cannot be ruled out with the current data. At the time of writing, the latest discovery about GRO~J1655--40 is the fact that magnetic fields are the only explanation for the launch of jets in this object by \citet{Miller-etal-2006}, who also use the distance of 3.2 kpc. The purpose of this paper is to complement the work of \citet{Foellmi-etal-2006b-astroph}, who have shown that the distance of GRO~J1655--40 must be smaller than 1.7~kpc, but focusing on how the distance of 3.2~kpc has been determined, and to show that this value is far from being firm. Other distance methods will be discussed in a forthcoming paper.


\section{The Radio-Jet Kinematic Distance of GRO~J1655--40}
\label{radio-distance}

\citet{Hjellming-Rupen-1995} first mention three references to say that GRO~J1655--40 lies at a distance of roughly 3~kpc: \citet{Harmon-etal-1995}, \citet{Tingay-etal-1995} and \citet{McKay-Kesteven-1994}, which are discussed below. Then, the authors present new radio data obtained with the \emph{Very Large Array} (VLA) and the \emph{Very Long Baseline Array} (VLBA). 

\subsection{The True Constraints from the Radio Data}

The 22 epochs of \emph{VLA} observations (see their Fig.~1) did not resolve the source at a level of 100 mas, but only a multi-cores object elongating with time. It is therefore necessary to use the more precise \emph{VLBA} observations. However, the authors emphasize clearly the lack of very-long-baseline interferometry calibrator, which implies that the data are self-calibrated, "eliminating all absolute positional information, and leaving the alignment of the different images a free parameter." Consequently, they must assume that the brightest point in each image is the stationary center of ejection. As mentioned in the paper, "these [VLA] and other [unspecified] data are consistent with constant intrinsic proper motion of 54 mas~d$^{-1}$" and later "the underlying proper motions appear constant". This is a very likely hypothesis, although it is only an hypothesis.

The authors mention that this hypothesis is strengthen by the fact that the (supposed) constant proper motions observed with the VLA and the VLBA agree. But what if this brightest point is not showing the true motion of the matter of the jets? Although very likely too, this hypothesis does not take into account the flux decays, and that the flux decay rate varies along the jet, as mentioned by the authors themselves. Moreover, the authors also mention the daily Southern Hemisphere VLBI Experiment (SHEVE) array observations of \citet{Tingay-etal-1995} which are consistent only with the major structures. The proper motion inferred from SHEVE data of 65$\pm$5 mas~d$^{-1}$ agrees with the 62 mas~d$^{-1}$ motion of the {\it outer edge} of the early NE ejecta. The relevance of choosing the brightest point of each VLBA image could then be questioned.

Finally, they build a kinematic model of the radio jets of GRO~J1655--40 ~based on these \emph{VLBA} observations. This is the method "C" of \citet{Jonker-Nelemans-2004}. They use the kinematic equation described in \cite{Mirabel-Rodriguez-1994} that link the apparent velocity of the receding and approaching radio jets ($\mu_+$ and $\mu_-$ respectively), and the distance $D$, the true jets velocity $\beta=v/c$ and the jet projection angle relative to the line of sight $\theta$:

\begin{equation}
\mu_{\pm} = \frac{\beta \sin(\theta)}{1 \pm \beta \cos(\theta)} \frac{c}{D}
\end{equation}

where $c$ is the speed of light. The problem with these relations is the number of known variables ($\mu_{\pm}$) and the number of unknowns ($\theta, D, v$). To find a distance, \citet{Hjellming-Rupen-1995} had to assume a value for the projection angle: $\theta = 85^{\circ}$. But a careful reading of this paper reveals that it is the maximum value allowed by the data, i.e. an upper limit only.

More precisely, the \emph{VLA} gives an estimation of the apparent true proper motion of the jets (assuming the two jets are moving apart with the same velocity) $v/D \sim 50$ mas~d$^{-1}$. It is then possible to rewrite the above equation as follows:

\begin{equation}
\mu_{\pm} = \frac{\sin(\theta)}{1 \pm \frac{v}{c} \cos(\theta)} \frac{v}{D}
\label{muplusmumoins2}
\end{equation}

and solving for $\theta$ by eliminating $v$. We obtain:

\begin{equation}
\sin \theta = \frac{1}{v/D} \, \left( \frac{2 \mu_+ \mu_-}{\mu_+ + \mu_-} \right) \leq 1
\label{eq4}
\end{equation}

From the measured values of $\mu_-$ and $\mu_+$ of 54 and 45 mas~d$^{-1}$ respectively with the \emph{VLBA} data, it means that $v/D \geq 49.1$ mas~d$^{-1}$, consistent with \emph{VLA} data.\footnote{The constraint on the inclination angle obtained by Hjellming \& Rupen is computed by eliminating $v/D$ in Eq.~\ref{muplusmumoins2}: $v/c = (\mu_- - \mu_+) (\mu_- + \mu_+)^{-1} \cos(\theta)^{-1} < 1$, which gives: $\theta \leq 84.8^{\circ}$.} Rewriting Eq.~\ref{eq4}, we obtain:

\begin{equation}
\frac{D}{c} \left( \frac{2 \mu_+ \mu_-}{\mu_+ + \mu_-} \right) \leq \frac{v}{c} < 1.
\end{equation}

where $c$ is the speed of light. Consequently, as noted by \citet{Mirabel-etal-2002}, only an upper limit to the distance can be obtained from the data: $D < c \, (\mu_+ + \mu_-)/(2 \mu_+ \mu_-) = 3.53 \; \textrm{kpc}$.

In summary, the data allow to derive an lower limit for the proper motion, and an upper limit for the inclination angle, and the distance. Quoting literally the paper of Hjellming \& Rupen: "For a distance of 3.2~kpc, this corresponds to $v \geq 0.91 c$, implying $84.3^{\circ} \leq \theta \leq 84.8^{\circ}$." Although not clearly stated, the reason why the authors chose 3.2 kpc is that this value is right in between their upper limit, and the lower limit given by {\it other} references.

\subsection{The True Constraints from the Quoted References}

As noted above, \citet{Hjellming-Rupen-1995} quote three other papers for a first estimation of the distance: \citet{Harmon-etal-1995}, \citet{Tingay-etal-1995} and \citet{McKay-Kesteven-1994}. Unfortunately, the first reference is quoting the two others for a distance value, and is therefore irrelevant for this issue. On the other hand, \citet{McKay-Kesteven-1994} is an IAU Circular in which it is simply stated that "HI observations of GRO~J1655--40 made with the \emph{AT Compact Array} show solid absorption in the velocity range $+10$ to $-30$ km~s$^{-1}$, with a further isolated weak feature at $-50$ km~s$^{-1}$. {\it The balance of probabilities is that the distance is around 3.5 kpc, unless the $-50$ km~s$^{-1}$ feature is due to an atypical cloud.}" This Circular is obviously not a measurement of the distance, and is based on the hypothesis that this feature at $-50$ km~s$^{-1}$ can be correctly interpreted as a HI cloud that is moving with the mean Galactic rotation. 

This is exactly what \citet{Tingay-etal-1995} also do, independently. Interestingly, the authors seem to {\it expect} a rather large distance in order to agree with a supposedly "significant reddening due to absorption", which is quoted from another IAU Circular: \citet{dellaValle-1994}. What this latter Circular states is that: "[...] The [optical] spectrum exhibits prominent, broad Balmer lines [...] superimposed on a relatively red continuum. [...]" However, the spectrum has been taken during an active state of the object, whose characteristics were, at that time, unknown, and the reddening is not at all measured. 

\citet{Tingay-etal-1995} present new \emph{VLBI} and \emph{ATCA} data of GRO~J1655--40. Their HI spectrum obtained with \emph{ATCA} (see their Fig.~2) shows a multi-component profile, with weak features at $-30$ and $-50$~km~s$^{-1}$ too. Attributing the absorption feature at $-30$~km~s$^{-1}$ to HI clouds participating in general Galactic rotation places a lower limit of 3.0 kpc on the distance. As noted by the authors, if the feature at $-50$~km~s$^{-1}$ was due to Galactic rotation, the implied minimum distance of GRO~J1655--40 would be of 4.2 kpc. But it couldn't be ruled out that this feature is due instead to HI driven by a $50$~km~s$^{-1}$ expanding shell or ionized material surrounding the adjacent Scorpius OB1 association, at a distance of 1.9 kpc \citep{Crawford-etal-1989}. They concluded that the minimum distance of GRO~J1655--40 is roughly around 3.0 kpc (obviously not providing uncertainties). Although it looks reasonable, this assumption might simply not be true. These measurements are very dependent on the distribution and velocities of various HI clouds along the line of sight; a difficulty that has been claimed to be important by \citet{Mirabel-etal-2002} who note that, in this direction, there are clouds with anomalous velocities up to $-50$ km~s$^{-1}$ \citep{Crawford-etal-1989}, i.e. with an amplitude similar to that of the weak feature observed in the \emph{ATCA} spectrum. 

The distance of 3.2$\pm$0.2 kpc has certainly been chosen by \citet{Hjellming-Rupen-1995} as a reasonable value between the lower limit of 3.0 kpc given by \citet{Tingay-etal-1995} based on a very weak assumption about moving HI clouds (although their distance range seems to be strengthened by other quick observational reports), and the upper limit provided by their radio data.

\section{Conclusions and Prospects}

It is important to stress that the exact and accurate distance of GRO~J1655--40 is still an open question. Before the direct measurement by the european satellite \emph{GAIA}, it should be possible to test the distance value using methods that have never been used so far. In particular, we want to mention a new (and exploratory) possibility using bisectors. 

First, \citet{Gray-2005} has shown that the bluemost point of single-line bisector is an indicator of the luminosity for late-type stars. This very interesting relation is weakened by the fact that high-quality single-line bisectors require a high spectroscopic resolution, and a very high Signal-to-Noise ratio (above, say, 300). However, the recent study by \citet{Dall-etal-2006} has shown that the bisector of the cross-correlation function (CCF) can be used as much the same way as single-line bisectors. The combination of these two results could lead to a first-order "direct" method of estimating the luminosity of the secondary star in GRO~J1655--40. Given the much larger brightness of GRO~J1655--40 in the near infrared (NIR, $m_K \sim 13$) than in the optical ($m_V \sim 17$), a CCF bisector could be obtained in a reasonable amount of time with the new near-IR echelle spectrograph on the VLT: CRIRES, which is being commissioned at the time of writing.  Such investigation is already underway. This new method might also proved to be useful for many other faint and embedded stars in star-forming regions.

\acknowledgements 
This work has been partly made in collaboration with T.H. Dall, E. Depagne and I.F. Mirabel. This research has made extensive use of NASA's Astrophysics Data System, of the ArXiv astro-ph, and the Central Bureau for Astronomical Telegrams of the IAU.


\end{document}